# Sharpening surface of magnetic paranematic droplets

Alexander Tokarev[1], Wah-Keat Lee[2], Igor Sevonkaev[3], Dan Goia[3], Konstantin G. Kornev [1,*]

[1]Department of Materials Science and Engineering, 161 Sirrine Hall, Clemson University, Clemson, SC 29634, USA

[2]Advanced Photon Source, Argonne National Laboratory, Argonne, IL 60439, USA

[3]Center for Advanced Materials Processing, Clarkson University, Potsdam, New York 13699, USA

[*]kkornev@clemson.edu

**Abstract**

**In a non-uniform magnetic field, the droplets of colloids of nickel nanorods and nanoparticles aggregate to form a cusp at the droplet surface not deforming the entire droplet shape. When the field is removed, nanorods diffuse away and cusp disappears. Spherical particles can form cusps in a similar way, but they stay aggregated after release of the field; finally, the aggregates settle down to the bottom of the drop. X-ray phase contrast imaging reveals that nanorods in the cusps stay parallel to each other without visible spatial order of their centers of mass. Formation of cusps can be explained with a model that includes magnetostatic and surface tension forces. The discovered possibility of controlled assembly and quenching of nanorod orientation under the cusped liquid surface offers vast opportunities for alignment of carbon nanotubes, nanowires and nanoscrolls, prior to spinning them into superstrong and multifunctional fibers. Magneto and electrostatic analogy suggests that similar ideal alignment can be achieved with the rod-like dipoles subject to a strong electric filed.**

## Introduction

With progress in nanotechnology, nanoparticles are finding new applications beyond their traditional use in paints, coatings, foods, drugs, and cosmetic products. Rod-like nanoparticles dispersed in a host fluid deserve a special attention because of their specific anisotropic interactions leading to transformation of these colloidal suspensions into colloidal liquid crystals with unique properties [1-4]. Traditional applications of colloidal liquid crystals are currently significantly extended to include precursors for multifunctional composites and fibers [5-14]. Ordering of rod-like nanoparticles inside the fiber precursor jet is a challenge: attractive van der Waals interactions among the particles lead to the particle clustering followed by separation from the host fluid. Stabilization of dispersions is achieved through physico-chemical functionalization of nanoparticles to counterbalance attractive forces by Coulombic, steric, or other repulsive interactions. The delicate balance between attractive and repulsive colloidal interaction can fail when the jet comes out from meniscus and other forces enter the game [7-9].

Here we report on the development of a physical principle of almost ideal alignment of magnetic nanorods under the free liquid surface. Due to magneto-static interactions between magnetic nanorods, they can be gathered on demand within a fraction of a second by applying a magnetic field gradient. Since the nanorod length is measured in tens of microns, only hundreds of nanorods are needed to form a millimeter long chain. Such long chains can be formed with millitesla magnetic fields (Fig. 1b) [13, 14] .The formation time of a micron size magnetic cluster can be made comparable or even shorter than that of a commercial printing device. Therefore, the suggested physical principle is of industrial importance and can be used for formation of magnetic fibers [12-14] and for printing magnetic droplets [15-18].

**Results and discussion**

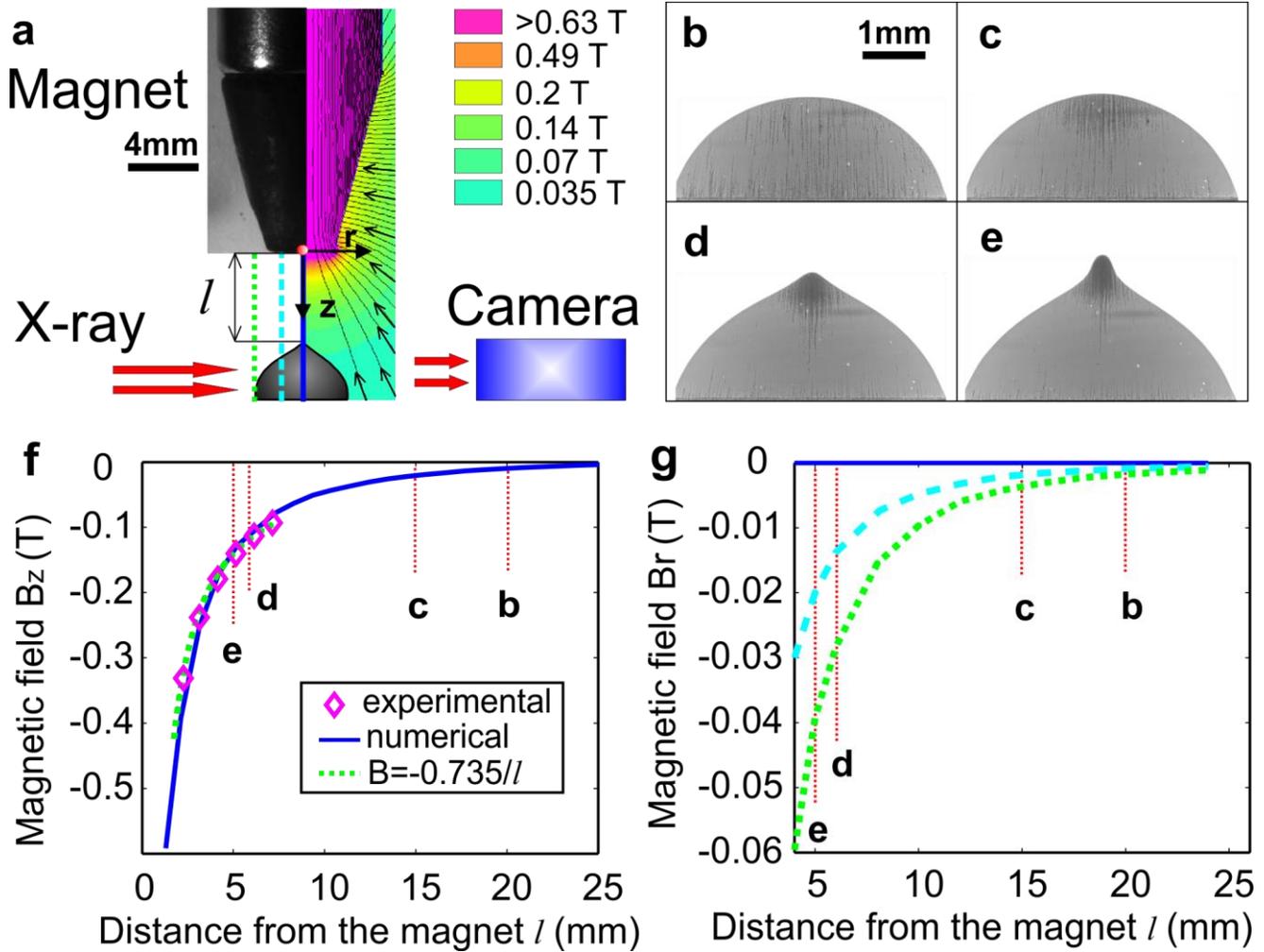

**Figure 1. a,** A schematic of experimental setup used in the X-ray phase contrast imaging experiment; the field distribution is specified by the magnetic flux lines calculated numerically with FEMM (http://www.femm.info) . **b, c, d, e**: X-ray phase-contrast images of the droplet at various distances $l$ from the magnet to the droplet. **f,** Axial magnetic field $B_z$ measured experimentally and calculated numerically with FEMM vs. distance $l$ from the magnet, point $l = 0$ corresponds to a central point at the magnet surface. **g,** Radial component of the magnetic field $B_r$ calculated along the vertical dashed lines shown in Fig. 1a.

A colloidal suspension of nickel nanorods of 200nm in diameter and 20 μm in length was prepared in ethylene glycol as described in Refs. [19], [20] and in the Materials and Methods section. A study of the nanorod clustering was performed using the X-ray phase contrast imaging at the Advanced Photon Source at Argonne National Laboratory, IL, with a beam energy of 33.2 keV. A schematic of the experimental setup is shown in Fig. 1a and a sequence of pictures illustrating the behavior of nanorods in magnetic field is presented in Fig. 1b-e. We moved magnet toward the droplet with a constant speed (V=0.2mm/s). Figures 1b-e show the droplet configuration and the structure of visible nanorod chains and clusters at different distances $l$ from the magnet (see Video S1). These visible chains and bundles (~300 μm long) are formed at a very weak millitesla range field. When the magnet is far away from the drop, the field is almost uniform and all nanorod bundles are distributed evenly over the droplet volume (Fig. 1b). Magnetic field orients the nanorods in the vertical direction, but the nanorods are free to move presumably forming a paranematic-type liquid crystal [2] where the crystal elasticity is mostly caused by the long range dipole-dipole interactions between nanorods [21-26]. As the magnet moves closer to the droplet, the nanorods form long chains from the droplet bottom to the top free surface. When the magnet is in a close proximity to the drop, producing a strong field gradient, these chains and strands come together to form a cluster concentrated at the droplet axis of symmetry near to the droplet surface. The explanation of the assembly of magnetic nanorods and chains is as follows.

Nickel nanorods behave as superparamagnetic nanoparticles showing almost zero magnetic moment in the absence of magnetic field and acquiring a magnetic moment linearly dependent on the field when the field is less than 0.1 T (Fig. S3). The magnetostatic energy of two magnetic dipoles oriented in the head-to-tail configuration is half of that observed in the configuration when the dipoles are placed side by side [22],[27]. Therefore, the distant nanorods have a tendency to form chains. Two parallel nanorods are prone not only to the head-to-tail ordering, but they have a tendency of sidewise attraction as well [28]. Therefore, the nanorods have a tendency to cluster in strands or bundles right after the application of magnetic field. When the nanorods are subject to a gradient field, one more mechanism of their clustering has to be considered. In a gradient field, the magnetic force acting on a nanorod is written as $\mathbf{F}_m = (\mathbf{m} \cdot \nabla)\mathbf{B} = -(m_z \partial/\partial z)\mathbf{B}$ where $m_z$ is a magnetic moment of the nanorod and **B** is the magnetic field vector. The "minus"- sign takes into account the fact that the magnetic moment is pointed in the negative direction of the z-axis. Therefore, the direction of the radial component of the magnetic force causing the nanorods to cluster at the central axis or spread away from it, depends on the z-derivatives of the $B_r$ - field component. This radial field component is directed toward the central axis, i.e. it is negative, and it fades away as the z-coordinate increases (Fig 1g). Accordingly, the radial component of magnetic force is negative pushing the nanorods to cluster at the central axis. As follows from Fig. 1f, the z-derivative of the vertical component of magnetic field is positive. Therefore, the z-component of magnetic force is negative: it tends to bring all nanorods to the top. These arguments explain the phenomenon of nanorod clustering at the top of the droplet closer to its central axis. The same arguments are applied to explain clustering of colloids of magnetic beads forming chains behaving like nanorods in magnetic field.

As the cluster size increases, it deforms the droplet by creating a sharp peak at the top (Fig. 1d). The rest of the droplet profile remains non-perturbed. The process of nanorod agglomeration with cusp formation in the droplets in Figs. 1 b), c), and d) is reversible: when the magnet is removed, the nanorods diffuse away and the droplets take on the original shapes. Remarkably, the entire droplet shape does not change during the cusping of its pole: the droplet deforms like a poppy head of the orthodox church.

When the magnet is moved to some critical distance, the cluster becomes a daughter droplet and jumps towards the magnet. Detachment of the daughter droplet cannot be stopped even if the magnet is moved away. Image in Fig. 1 e) shows the drop profile when the cluster is about to leave the drop.

This behavior of the nanorod-laden droplets is very unusual and has never been documented in the literature on colloids of magnetic nanobeads, known as ferrofluids [27, 29-31]. The published experimental observations on ferrofluid droplets deal with a uniform magnetic field [29-32]. In this case, the entire droplet deforms and this behavior is very much similar to the behavior of dielectric droplets exposed to an electric field [33-35] with some specific features attributed to a nonlinear relation between droplet magnetization and applied field [31].

We studied the behavior of droplets from Ni nanobeads in a gradient field. Different diameters of nanoparticles have been used in these experiments: 30 nm, 165 nm and 605 nm, respectively. The concentration of nanobeads was kept the same as that in the experiments with Ni nanorods of 200 nm diameter. Similar poppy heads were obtained even for the smallest studied concentrations of nanoparticles (Fig. 2). The main difference between nanorods and nanobeads documented experimentally is that when the magnetic field is removed, nanobeads sink down while maintaining the cluster structure whereas in the nanorod case, nanorods stay at the drop interface and diffuse away (Fig. S7 and Fig. S8).

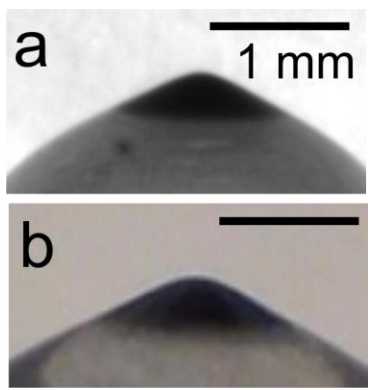

**Figure 2.** Conical cusps of droplets with **a,** nanorods (200 nm in diameter, 20 μm in length, 0.07 wt %) and **b,** nanobeads (100nm in diameter, 0.07 wt %).

This difference in nanoparticle clustering can be explained by a significant difference in the field distribution around uniformly magnetized nanorods and beads. Long nanorods interact through their poles, i.e. through their "magnetic charges" with the energy of two poles inversely proportional to the distance between them [36]. Short nanorods interact as nanobeads, i.e. as dipoles [27]. As follows from the phase diagram for two nanorods [37], the greater the nanorod length-to-diameter ratio, the greater the zone of their repulsion. Therefore the aligned nanorods at the outmost interfacial layer are held together mostly by the field gradient: when the field is released, the sidewise repulsion of the nanorods forces the cluster to break apart and the thermal excitations help nanorods to diffuse away [37, 38]. The phase diagram for two nanobeads reveals a larger area of the attraction zone: where the nanorods of finite length repel each other, the nanobeads of the same magnetization can come together [37]. Therefore, clusters of magnetic beads is very difficult to destroy [39-43]. As revealed by the SEM images in supplementary material, magnetic beads are facetted. Therefore, in addition to the strong dipole-dipole interactions, these beads are expected to experience strong van der Waals interactions keeping them together.

Except of the structure of nanoparticle packing in the cluster, the behavior of poppy heads for different colloids is very much similar. Therefore, we hypothesized that only capillary, gravitational, and magnetic forces are crucial for cusping the droplets. Fig. 1e shows the first irreversible shape of the nanorod laden droplet at critical distance $l_{cr}$. At this moment, the whole cluster filled with magnetic nanorods detaches from the mother droplet and jumps toward the magnet. This is a manifestation of an interfacial instability when the surface tension is unable to support the surface deformations and any infinitesimal perturbation of the surface results in an eruption of the droplet interior [44]. A similar behavior of magnetic cusps was observed with nanobead droplets. Therefore, we assume that the structural elasticity of magnetic aggregate plays no role and the cluster acts like a solid magnetic body pulling the droplet surface toward the magnet. Estimation of the Bond number, Bo = $\rho g R^2 / \sigma$, revealed that it is less than 1, hence gravity is insignificant in these experiments. Thus the drop shape is controlled mostly by the capillary forces; these forces counterbalance the magnetic force pulling the droplet at the cusp. Therefore, the droplet profile excluding the cusped pole is expected to be similar to an equilibrium unduloid describing the shape of a droplet sitting on a wire, Fig. 3c [45]. In the drop-on-a-wire model, the unduloid is pulled in the opposite directions by the interfacial tension of the wire surface acting at the contact lines.

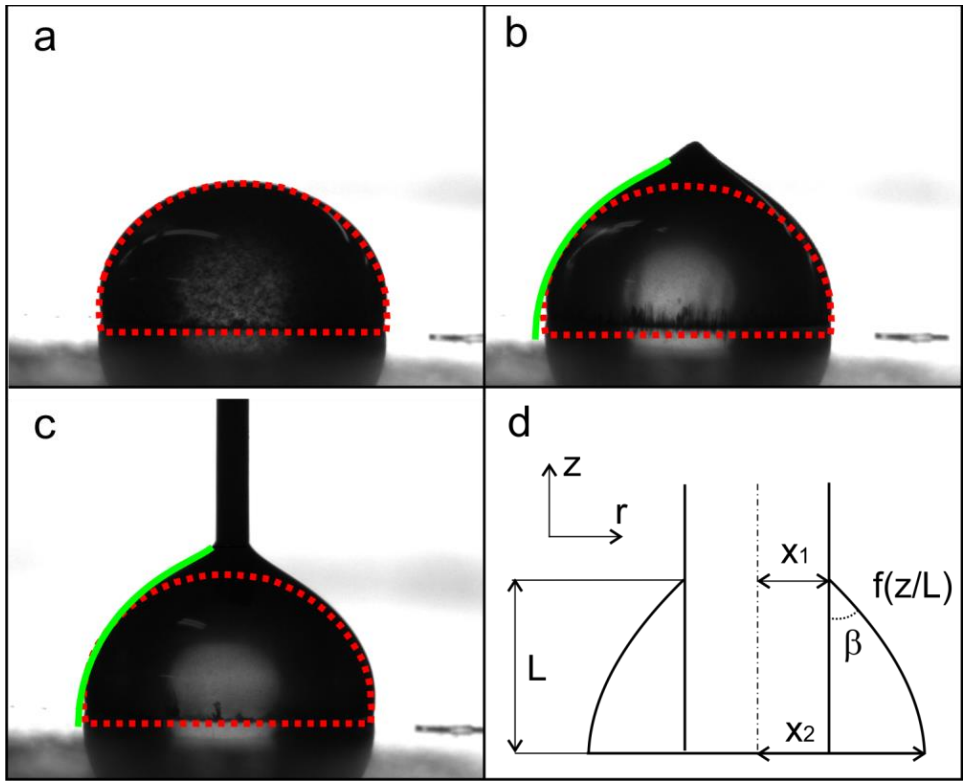

**Figure 3. a,** Droplet with nanorods dispersion sitting on a glass slide. **b,** Shape of the nanorod droplet at the critical distance *l*cr from the magnet. **c,** Stainless steel wire with diameter d=0.46mm is inserted in the same droplet. **d,** Model parameters used to fit the droplet profile with an unduloid.

The unduloid corresponding to the given droplet is a solution of the Laplace equation of capillarity [45]; it is defined by a function $f(z, x1/L, x2/L, \beta)$, the radial coordinate of the droplet profile at point z along the wire; *x1* is the wire radius, *x2* is the droplet radius, *L* is the droplet length, and $\beta$ is the contact angle (see the definitions in Fig. 3d and Supporting Information for the analysis). Adjusting the contact angle $\beta$ and the droplet height *L*, one can fit the droplet shape.

In order to confirm this analogy, the magnet was removed and different wires with diameters in the range between 0.07 mm < d < 1.5 mm were brought in contact with the droplet. We examined copper wires with diameters d=0.4, 0.8, 0.85, 1 and 1.5 mm; tungsten wires with d=0.127 and 0.07 mm, and stainless steel wires with d=1, 0.8 and 0.46 mm. Only upon immersion of a stainless steel wire with d=0.46 mm, we observed an unduloid, the solid line in Fig. 3c, which was identical to that shown in Fig 3b corresponding to the droplet shape at the onset of instability. Seven droplets with different nanorod concentrations were analyzed to obtain an apparent contact angle using Carroll's solution. In all cases we found that the angle was constant $\beta = 49°\pm0.7°$ (n =7 droplets).

This independency of the apparent contact angle on nanorod concentration and hence on cusp magnetization is surprising and it deserves a special attention. As shown in the Supporting Information, inside the cusps with high magnetic permeability, magnetostatic potential is almost constant. Therefore the description of the distribution of magnetic field outside the cusp follows an analogous description of the distribution of electric field outside conical conductors[36, 46]. The external magnetic field generates a specific distribution of magnetic poles of the same polarity at the cone-like liquid cusp. These poles repel each other and want to destroy the cone, but the capillary forces push the free surface back. This subtle competition results in a unique conical cusp with the angle $\beta = 49.3°$. These liquid cones are identical to the famous Taylor cones observed on water drops and soap films, i.e. conductors [33, 46]. In this model, the specificity of the nanorods is not important: the same cone-like cusp was observed on the nanobead droplets, hence the model is applicable to that case as well. We discuss this electrostatic analogy in the Supporting Information.

Using the droplet-on-a-wire idea, one can say that the conical cusp acts as a wire pulling the droplet towards the magnet. We further confirm this hypothesis by analyzing the forces acting on the magnetic cusp. The free body diagram shown in Fig. 4a specifies these forces: magnetic forces pulling the cone toward the magnet, capillary pressure acting at the base of the magnetic cone, and surface tension forces acting at the edge of magnetic cone.

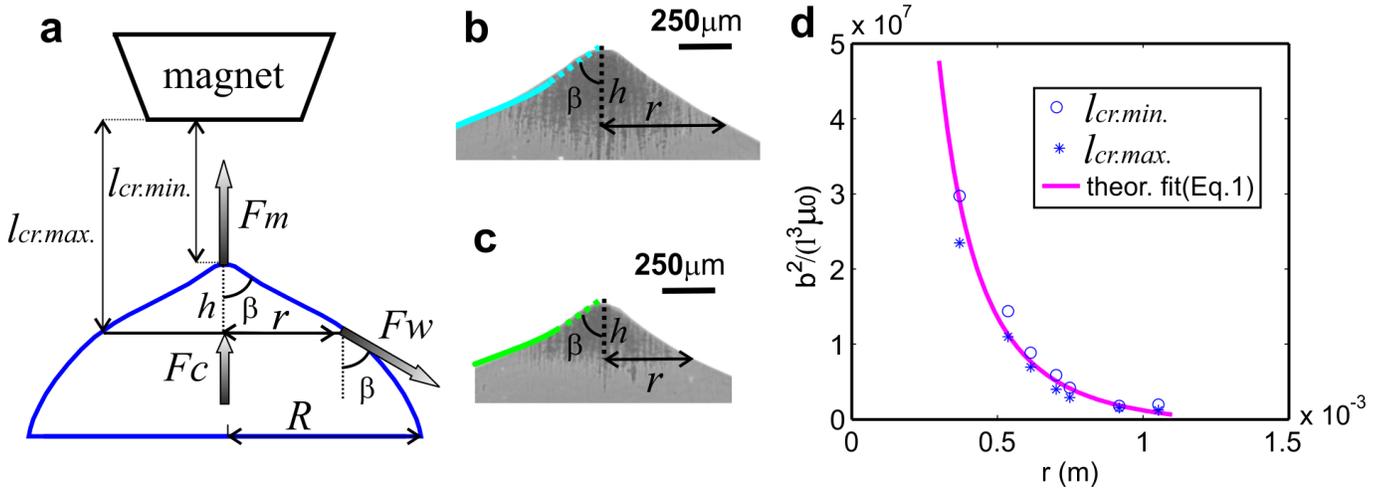

**Figure 4. a,** Forces acting on the cluster of magnetic nanorods at critical distance $l_{cr}$ . **b, c** Cluster formed by magnetic nanorods at the top of the droplet for 2 different concentrations $c$ of nanorods in the initial droplet $c(\mathbf{b}) > c(\mathbf{c})$, resulting in different cluster radii $r$. **d,** The solid line shows the theoretical prediction and the circles represent the calculated values of the left side in equation (1) at $l_{cr.min.}$, the stars represent the calculated values of the left side in equation (1) at $l_{cr.max.}$ Concentrations of nanorods in the droplets corresponding to the open circles are: from right to the left, %wt - 0.29, 0.24, 0.20, 0.18, 0.13, 0.10, 0.07.

The magnetic force exerted by the magnet is written as $F_m = (\mathbf{m} \cdot \nabla)\mathbf{B}$, where $\mathbf{m}$ is the magnetic moment of the cone. As shown in the Supporting Information, the nanorod magnetization in an applied magnetic field follows the linear constitutive equation, $\mathbf{m} \propto \mathbf{B}$. The measured and FEMM − calculated magnetic field can be approximated as $B_z = b/l = -0.735/l$ (the dashed line in Fig. 1f). With this approximation, magnetic force acting on the magnetic cone can be written as $F_m = \alpha b^2 V / l^3 \mu_0$ where $\mu_0$ is magnetic permeability of vacuum, $V$ is the volume of magnetic cone, and $\alpha$ is a constant to be determined.

The surface tension $\sigma$ acting at the edge of magnetic cone of radius $r$ results in the tensile force with the z-component $F_w = -2\pi r \sigma \cos(\beta)$. The rest of the droplet contributes to the free body diagram through the capillary pressure distributed over the bottom of the magnetic cone. This pressure results in the capillary force $F_c = 2\sigma \pi r^2 / R$ where $R$ is a radius of the mother droplet. Collecting all forces, the force balance is represented as

$$\frac{b^2}{l^3 \mu_0} = \frac{6\sigma}{\alpha \tan \beta}(\frac{\cos \beta}{r^2} - \frac{1}{rR}) \tag{1}$$

In order to find constant $\alpha$, we conducted experiments with 7 droplets of the same size (R = 2.23±0.2mm) and with different initial concentrations of nanorods in the droplet (in the range from 0.043wt% to 0.4wt%). The cone radius $r$ shrunk as the concentration of nanorods in the initial droplets decreased (Fig. 4b,c). This decrease of the cusp size required to bring the magnet closer to it to detach the cusp. During these experiments, we measured the cone radius $r$, the distance from the peak of the droplet to the magnet, *l*cr.min., and the distance from the base of the cone to the magnet, *l*max.. The results of these experiments are shown in Fig. 4d. For each given cluster radius $r$, the circles correspond to the left hand side of equation (1) taken at *l*cr.min. The stars correspond to the left hand side of equation (1) taken at *l*cr.max. The solid line is the theoretical fit with function $f(r) = \frac{6\sigma}{\alpha \tan \beta}(\frac{\cos \beta}{r^2} - \frac{1}{rR})$, where $\alpha$ was used as an adjustable parameter. The best fit was reached for $\alpha = 0.018$.

This analysis reveals an interesting universality of the magnetic cusps. It appears that magnetic cusp can be considered as a "frozen magnetic cluster" acting on the drop as a solid object. The unduloid solution serves as an asymptotic solution of the Laplace equation of capillarity for a mother droplet. The contact angle where the unduloid meets the magnetic cusp is well defined and it is universal for all droplets of different generations. It is equal to $\beta = 49^0$, the famous Taylor angle [46]. These experiments and model of cusp formation where the magnetostatic and surface tension forces control the cusp shape show an interesting possibility to align nanorods in a meniscus by applying a magnetic field. Magnetic and electrostatic analogy suggests that similar ideal ordering can be achieved with the rod-like dipoles subject to a strong electric filed. This possibility offers vast opportunities for preodering of carbon nanotubes, nanowires and nanoscrolls prior spinning them into superstrong and multifunctional fibers.

**Materials and Methods**

Nickel nanorods (200nm x 20μm) were produced by electrodeposition of Ni inside the pores of alumina membranes [47]. Ethylene glycol was added to the beaker with nickel nanorods and after sonication for 30 seconds (Branson Sonifier 450), a uniform suspension of Ni nanorods was produced. Spherical single crystal and highly crystalline uniform Ni nanoparticles of different sizes were obtained via reduction of the nickel carbonate basic salt (The Shepherd Chemical Company) in high quality grade di-ethylene glycol (99.99% DEG). The polyol served as a reducing and dispersing agent, and as a medium. Final size of nickel particles was controlled by the limited seeding mechanism. Based on that method, dispersions with the particles of diameter d=30±6.1, 165±27.2, and 605±69.5 nm were produced (see Figure S4 and SI for details on synthesis). A double sided tape (MMM237, 3M) was attached to a glass slide to prevent the droplet spreading. A permanent cylindrical magnet (K&J Magnetics, grade ND42) with a tapered tip was attached to the XYZ linear stage (VT-21, Mikos) to allow the alternation of the field gradient by moving the stage with a 100 nm minimum step in the vertical direction. The process was recorded with a Photron FastCam and Dalsa cameras at 100 fps. Videos were analyzed with VirtualDub (http://www.virtualdub.org) and ImageJ (NIH) softwares. Magnetic field was calculated numerically by using the FEMM software (http://www.femm.info) (Fig. 1a) and then the numerical results were confirmed by the measurements of magnetic field at the central axis with a digital teslameter (133-DG GMW Inc.) positioned with the manipulator (VT-21, Mikos) at different distances $l$ from the magnet (diamonds in Fig. 1f). For further analysis, we approximated the axial component of magnetic field by formula B=$b/l$=-0.735/$l$ which was proven correct in the range 1.7 mm < $l$ < 7.1mm (the dashed line in Fig. 1f).

**Acknowledgements**

The authors are grateful for the financial support of the National Science Foundation through Grant EFRI 0937985, and of the Air Force Office of Scientific Research through Grant FA9550-12-1-0459. We also acknowledge Sigma Xi Grants-in-Aid of Research G20100315153485 and G20100315153500. Use of the Advanced Photon Source, an Office of Science User Facility operated for the U.S. Department of Energy (DOE) Office of Science by Argonne National Laboratory, was supported by the U.S. DOE under Contract No. DE-AC02-06CH11357.